# Integrating Magnetism into Semiconductor Electronics

V.L. Korenev[*] and B.P. Zakharchenya

*A.F. Ioffe Physical Technical Institute, Russian Academy of Sciences, 194021, Saint-Petersburg, Russia*

**Vision of ferromagnet/semiconductor hybrid as a strongly coupled but flexible spin system is presented. We analyze the experiments and argue that contrary to the "common sense" the nonmagnetic semiconductor plays a crucial role in manipulating of ferromagnetism. The magnetism of the hybrid (magnetic hysteresis loop and the orientation of magnetization vector in space) is tuned both optically and electrically with the help of semiconductor. As a result the hybrid represents an elementary magnetic storage with electronic record and readout.**

*Corresponding author: *korenev@orient.ioffe.rssi.ru*

**1. Introduction**

Operation of the most of electronic devices exploits electron flow control by electric and magnetic fields (hence the name "electronics"). Besides charge the electron possesses spin – the internal angular momentum, or using the language of classical physics, rotates about itself. Spin, however, is of a quantum nature having only two projections "up" and "down" at a fixed direction. Computers use a large number of electrons with ferromagnetically (parallel) ordered spins to store information. The total spin looking up (down) corresponds to "one" ("zero"). The part of electronics using spin besides charge has been recently called – " spintronics" [1]. The first application of spintronics is based on giant magnetoresistance effect discovered by A. Fert group in 1988 [2]: the resistance of three-layered metallic structure ferromagnet/paramagnet/ferromagnet depends on the mutual orientation of magnetic moments of adjacent ferromagnets. This effect underlies the operation of readout heads in the modern hard drives [3].

Nowadays large efforts are directed to the integration of magnetism into semiconductor architecture of modern computers. It may give a possibility to create the whole computer on a single semiconductor chip without hard drives, i.e. with purely electronic record/readout of magnetic information. For this purpose it is desirable to create elementary magnetic storage with electronic access. One approach is to construct a universal object combining ferromagnetic (FM) and semiconducting (SC) properties. Ferromagnetic semiconductors (Eu-based chalcogenides and Cd-Cr spinels) were studied extensively in the 60-s of the last century (Ref. [4] for review). Their main disadvantages are the low Curie temperature $T_c$ and incompatibility with standard semiconductor technology based on Si, Ge, GaAs. The recent years have been devoted to the active search [5, 6] for other ferromagnetic semiconductors. The main obstacle in this direction is the necessity to satisfy a whole series of demands within a single structure.

Alternative approach is oriented on the ferromagnet/semiconductor (FM/SC) hybrid systems [7]. One of the advantages is an additional degree of freedom consisting in the choice of desirable pair FM/SC among paramagnet semiconductors and a large number of ferromagnetic materials. Another advantage is the possibility to read magnetization of FM with the help of semiconductor, for example, using the Hall effect or electrical injection of spin polarized electrons from FM into SC in a three-layered structure FM/SC/FM. In the latter case the semiconductor electrons acquire non-equilibrium polarization monitored by the second FM. However, semiconductor is "passive": it plays either a pitiful role of a substrate or only reads FM magnetization.

At first glance the influence of SC on magnetic properties of FM is negligible because the density of charge (and spin) carriers in non-magnetic (paramagnet) SC is much less than the density of magnetic atoms in typical ferromagnets. Therefore, "common sense" suggests that a semiconductor is not capable to manipulate the powerful ferromagnetic spin system. However, the contact of FM film and SC leads to the band bending in SC (Schottky barrier) and accumulation of a fair quantity of charge carriers (electrons or holes) near interface. The strong exchange interaction (Coulomb interaction restricted by the Pauli exclusion principle) of SC charge carriers and magnetic atoms of FM leads to the unified spin system whose properties differ substantially from those of an isolated FM film. The uniqueness of the FM/SC system lies in the electrical and optical tunability of electrical and magnetic properties of paramagnet SC. This means that the ferromagnetism (for example, magnetic hysteresis loop) of the unified system can be tuned as well. As a result, SC is not only a substrate for a FM to lie upon but participates *actively* in information processing. It both controls and monitors ferromagnetism. The effect of semiconductor on ferromagnet is larger for thin FM films. Nanoscale films are used actively for the dense writing of information. As a result the FM/SC hybrid looks like a promising elementary magnetic storage with electronic control.

Here we consider the current state of research on the unified FM/SC system and show that contrary to the "common sense" semiconductor charge carriers actively participate in information processing both optically and electrically.

## 2. Optical readout and control of ferromagnetism in FM/SC hybrid

The first example of semiconductor-assisted optical monitoring and control of magnetism was presented 10 years ago [8, 9] in a nickel-gallium arsenide (Ni/GaAs) hybrid. The non-equilibrium spin of optically oriented semiconductor electrons was used for the readout purpose. The classical scheme of optical orientation [10] consists of generation of spin-polarized electrons by circularly polarized light transferring the photon angular momentum into the spin system of semiconductor electrons. In its turn, annihilation of the electrons emits circularly polarized light. The degree ρ of the circular polarization of photoluminescence (PL) is equal to the projection of electron ensemble-averaged spin **S** onto registration direction z usually coinciding with the normal to the structure plane (Fig.1). An external magnetic field **H** induces rotation (precession) of each electron spin about **H** with Larmor frequency ω=γH (γ is gyromagnetic ratio). In steady state conditions, however, the average spin **S** does not change with time but deflects from the initial direction (z) decreasing in absolute value (Fig.1). This leads to the Hanle effect – depolarization of PL with magnetic field. The halfwidth $H_{1/2}$ of the magnetic depolarization curve is determined by the equality of frequency $γH_{1/2}$ and reciprocal non-equilibrium spin lifetime $1/T_s$. The longer the spin lifetime $T_s$, the smaller magnetic field necessary to rotate spin and depolarize PL. As a result, there is the Hanle effect-based optical magnetometer whose sensitivity is determined by the $H_{1/2}$ value. The specific $H_{1/2}$ value depends on the state of matter. For example, optical orientation of paramagnet atoms of gas vapors is conserved during record long time $T_s$~1 s resulting in very low values of the halfwidth $H_{1/2}$ (down to 1 μOe [11, 12]). The spin lifetime shortens drastically in condensed matter due to the enhancement of non-spin-conserving interactions.

The smallest values of the halfwidth in GaAs-type semiconductors are realized in n-GaAs samples [1] ($H_{1/2}$~1-10 Oe corresponding to spin lifetime $T_s$~$10^{-7}$-$10^{-8}$ s). To trace the state of the ferromagnetism in the Ni/GaAs hybrid we used [9] n-type gallium arsenide where at low temperature (T=4.2 K) the halfwidth $H_{1/2}$ = 2 Oe corresponding to $T_s$≈130 ns. Stray fields near the FM film surface are the strongest when the film is demagnetized, i.e. broken into a large number of domains with different magnetization **M** orientations. Precession in static stray fields dephases electronic spins in nearby SC and decreases the average spin of electrons and PL polarization. Figure 2a compares the Hanle effects in GaAs when the film is previously magnetized (blue color) and demagnetized (red color). It is seen that the demagnetization decreases the degree $\rho$ with the maximum difference being achieved in zero magnetic field.

---

[1] The long-term spin memory of electrons in n-GaAs was discovered in the 70-s ($T_s$≈25 ns, $H_{1/2}$=10 Oe, C. Weisbuch, *PhD Thesis*, Paris, 1977). It results from the weak spin-orbit interaction in conduction band and the absence of holes providing an effective source of spin relaxation [10]. The macroscopically long spin diffusion length $L_s$>10 μ in n-GaAs was found by our group in 1994 but published at a later date in R.I. Dzhioev, B.P. Zakharchenya, V.L. Korenev and M.N. Stepanova, *Phys. Solid State* **39** 1765 (1997). We have exploited this knowledge in 1995 in Ref. [9] where n-GaAs with $H_{1/2}$=2 Oe ($T_s$=130 ns), $L_s$=13 μ has been chosen. This enabled us to detect a few nanometers thick ferromagnetic interface NiGaAs whose stray fields (~1 Oe) penetrate deep into SC (see the main text).

The magnetic field strength $h_c$ making the magnetic moment of the sample vanish is referred to as the coercive force. It can be estimated by measuring the zero-field polarization $\rho(H=0)$ after switching of a previously magnetized FM film by external magnetic field of strength H*. When the switching field H* is equal to the field $h_c$ the film is demagnetized and the polarization value $\rho(H=0)$ is minimal. Data points on the upper dependence in Fig.2b were measured after switching in the *dark*. The sharp minimum corresponds to $h_c$=90 Oe. Remarkably, that the coercive force value appeared to be 2.5 times larger than that measured in the same structure with the use of a superconductor quantum interference device (SQUID). The difference between the two techniques is that SQUID registers the total magnetic moment of the structure. It is mainly the magnetic moment of the nickel film whose thickness (40 nm) is much larger than that of the interface NiGaAs (a few nm). However, the contribution of stray fields of nickel and the interface into the Hanle effect is not reduced to the sum of their magnetic moments. Indeed, the non-equilibrium electron spin diffuses into n-GaAs over the distance $L_s \approx 13$ μ [9] that is 10 times longer than the 1 μ size of nickel film domain [13]. Therefore the nickel stray fields decay quickly near the surface and the basic mass of electrons does not "feel" them. On the contrary, the interface stray fields penetrate deep ($\geq L_s$ [9]) into SC and dephase electron spins inside. Therefore, the Hanle effect magnetometer detects the ferromagnetic interface NiGaAs rather than the Ni film itself due to the space selection of their stray fields.

The important feature of the Ni/n-GaAs hybrid consists in the optical tunability of the *interface* ferromagnetism. The illumination of the sample decreases the interface coercivity by half (compare the lower dependence on Fig.2b with the upper one) but does not affect the coercive force of nickel film. The effect we called photocoercivity is not sensitive to the light polarization and takes place only in the illuminated region. This allowed us to make optical record/readout cycle on the interface NiGaAs. An external switching field equal to the coercive force on the light (45 Oe) was applied to the previously magnetized Ni/GaAs sample. It cannot affect magnetization in the dark. However, local shining by He-Ne laser demagnetized the illuminated parts, which led to the optical record: magnetized regions corresponded to "one" whereas demagnetized – to "zero". Optical readout was done with the use of spin-polarized SC electrons: electron spin **S** value was smaller (larger) under the demagnetized (magnetized) regions. Photocoercivity takes place at a low power density (~10 mW/cm$^2$) and is not connected with heating of sample [2]. Thereby this approach differs drastically from the standard thermal record when the FM is heated by light up to Curie temperature. Spectral measurements have shown that the photocoercivity is due to *the effect of semiconductor on ferromagnet*: it diminishes if the photon energy (hv) is below the energy gap ($E_g$) of GaAs. A possible explanation [14] of photocoercivity is based on the optical control of exchange interaction of SC electrons, accumulated near interface, with nearby magnetic atoms.

Today there are experimental data confirming the active role of SC in the optical readout [15] of magnetization and the optical control [16] of coercive force. In the former case the SC photoelectrons acquire the non-equilibrium polarization which is proportional to **M** as a result of spin-dependent transport through the FM/SC junction [17]. In the latter case the photo-excitation of InMnAs/GaSb hybrid changes the coercive force of InMnAs FM film. The authors [16] have found a significant decrease of the coercive force under illumination by light with energy quanta being larger than the energy gap of nonmagnetic semiconductor GaSb. The photocoercivity decreases when hv gets closer to the energy gap pointing to the important role of SC.

It is worthwhile to note the paper [18] on the photo-induced ferromagnetic order in InMnAs/GaSb hybrid. Excitation with photon energy quanta larger than the energy gap of GaSb induces the paramagnet-ferromagnet transition of magnetic semiconductor InMnAs. This can be interpreted as the light-induced Curie temperature enhancement. Such an effect is known for ferromagnetic semiconductors [4] but in this case it is induced with the help of nearby non-magnetic semiconductor.

---

[2] We heated the hybrid by passing the electric current across the FM/SC heterojunction [9] with dissipated power being 10 times larger then the power of light. However, the coercive force was unchanged.

A different kind of phenomena, found recently in the GaMnAs/GaAs [19], is the magnetization of GaMnAs FM film induced by a circularly polarized light. Authors explain it by the photocreation inside of the GaMnAs of spin-oriented holes that dynamically polarize the Mn spins. However, very sharp spectral dependence correlates with the excitation of paramagnetic GaAs rather than ferromagnetic GaMnAs whose absorption edge is shifted to higher energy even at small Mn concentrations ~1% [20] as a result of Mn doping. This fact provides a strong evidence of the crucial role of GaAs excitation in the optical magnetization of GaMnAs.

Theoretically this effect may be related with the exchange interaction of GaMnAs magnetic atoms and optically oriented holes in the GaAs accumulation layer near the interface FM/SC [21]. In this case FM undergoes the action of the effective exchange magnetic field $\mathbf{H}_{eff}$ whose direction is determined by the helicity of circularly polarized light [3] [22].

## 3. Electrical readout and control of magnetism of FM/SC hybrids

Integration of ferromagnetism into semiconductor electronics requires information to be manipulated not so much optically as electrically. Two basic approaches to the SC-assisted electrical readout can be emphasized. The first one [1] is based on the dependence of the resistance across FM/SC/FM structure on the angle between magnetizations of two ferromagnetic layers separated by a semiconductor. It is the result of different tunnel barrier weights for "up" and "down" spins. Also it is possible due to accumulation of non-equilibrium spin in SC under electrical spin injection through the contact FM/SC [23]. The second approach provides readout even in two-layered structures FM/SC. It is based on spin-orbit interaction whose origin is as follows. The spin of an electron moving in an electric field undergoes the action of the effective magnetic field whose value and direction are determined by those of both the electron velocity and the electric field. As a result of spin-orbit coupling the conductivity of FM/SC hybrid layer will depend on mutual orientation of electric current and magnetic moment. For example the anomalous Hall effect (dependence of Hall coefficient on **M**) is widely used [24] to detect spin polarization in ferromagnets and non-magnetic semiconductors.

---

[3] Note that, such a field will shift the magnetic hysteresis loop by the $H_{eff}$ value (unidirectional or exchange anisotropy [21]). If it is larger than the coercive force $h_c$, the complete magnetization by circularly polarized light is possible even in the absence of external field.

An important step toward the SC-based magnetic storage with electronic access consists in the SC-mediated electric control of ferromagnetism. The magnetization vector of the uniformly magnetized FM film can be rotated electrically with the help of paramagnet semiconductor as a result of FM/SC exchange coupling [25]. Consider the exchange interaction of FM atoms and hole gas in a nearby semiconductor quantum well (Fig. 3) in the absence of external magnetic field. The quantum well consists of a thin layer of narrow-band SC restricted from both sides by wideband materials. They serve as barriers confining the motion of charge carrier (size quantization) along the growth direction by the region thinner than the length of de Broglie wave. Selective doping of the lower barrier with acceptors (dashes on Fig. 3) extracts electrons from the well leaving the holes (open circles on Fig. 3) behind. The hole states in the valence band $\Gamma_8$ of GaAs-type semiconductors are described by angular momentum 3/2. The strong spin-orbit coupling combined with size quantization split the valence band. The ground state corresponds to the so-called heavy holes whose spin is pinned to the normal to the well plane. In this case [26] the FM/SC exchange interaction is anisotropic: the FM/SC exchange energy $E_{exc} = -Jnm_zS_z$ (per unit area) is proportional to the product of $S_z$ component of an average hole spin (with surface density n) and *z*-component of unit vector **m** along **M**. The exchange interaction splits the "up" and "down" heavy hole states by the value J~0.1 eV inducing their high spin polarization $S_z$~1. In their turn, polarized holes create an exchange field **H$_h$** forcing magnetization along the normal and so decreasing the energy $E_{exc}$ (Fig.3). It can be determined by equating the $E_{exc}$ to the magnetic potential energy $-M_zH_hd$ due to the field: $H_h = JnS_z/Md$ (d is the thickness of FM film).

However the deviation of **M** out of the plane increases magnetostatic energy due to magnetic dipole interaction. It is followed by appearance of demagnetization field $H_D = -4\pi M_z$ ($H_D \sim 1$ kOe) [27] because the normal component $B_z = H + 4\pi M_z = 0$ of magnetic induction **B** is continuous at FM/SC interface. The exchange and demagnetization magnetic fields are equal in equilibrium conditions [4]. This enables us to estimate the ratio between the surface densities of holes and magnetic atoms $N \approx Md/\mu_B$ so as to manipulate the direction of **M**: $n/N \approx \mu_B H_D/J \approx 10^{-4}$. In other words, one hole is able to control ten thousands (!) magnetic atoms. The origin of this prima facie surprising result is that the anisotropic exchange interaction of FM and SC overwhelms rather weak magnetic dipole interaction. Note, that the giant exchange energy of FM itself is not affected [5] because the spins of magnetic atoms rotate parallel to each other: the isotropic exchange interaction inside FM fixes the **M** value but not direction. Unlike FM, the strong spin-orbit coupling in valence band of SC together with size quantization fixes the direction of hole spin orientation leading to the anisotropy of FM/SC exchange coupling. In equilibrium the spin-polarized holes induce in FM the easy magnetization axis along the normal. In this case the coupling energy $E_{exc}$ can be considered as the magnetic anisotropy energy. It can be governed by applying the bias to the gate (up brown layer on Fig.3) through the change of hole concentration or the overlap of hole wavefunction and magnetic layer, i.e. value of J.

The exchange coupling of MnAs ferromagnetic layer and holes in the nearby GaAs quantum well has been observed [28] by measuring circular polarization of PL from the quantum well. It has been found that exchange interaction does induce the equilibrium polarization of holes (about 10 %) that can be controlled electrically. The electric control of ferromagnetism was realized previously [29] in ferromagnetic InMnAs grown on a paramagnetic InAs semiconductor nanostructure. Magnetic hysteresis loop of InMnAs was changed by applying the bias ($V_G$) to the structure gate on top of FM. Cardinal reconstruction of the loop with bias was related [29] with a small change of Curie temperature $T_c(V_G)$ inducing the paramagnet-ferromagnet transition. However, further experiments are necessary to justify this statement [6]. A recent paper [25] proposed another explanation based on the exchange interaction of FM with holes of nonmagnetic 5 nm-thick InAs layer beneath. In this case the structure InMnAs/InAs should be considered as strongly coupled hybrid system. The negative bias accumulates holes in InAs semiconductor nanolayer inducing the easy magnetization axis perpendicular to the structure plane (perpendicular anisotropy). This leads to hysteresis behavior in the external magnetic field perpendicular to the plane. Depletion of InAs layer at positive bias puts the vector **M** in plane, and the magnetization by external field reveals non-hysteretic behavior. The fact that the removal of the InAs layer leads to non-hysteretic magnetization [30] also points in favor of this explanation. To check such an explanation, it is necessary to measure the complete orientation of **M** (not only its z-component [29]).

---

[4] Here we consider the spin-orbit effects in the ferromagnet to be weak. They are necessary only *to fix* the **M** direction in the plane of the film (along *x* axis). In contrast to it, the original paper [25] assumed the uniaxial anisotropy field coming from spin-orbit interaction to be stronger than the demagnetizing field.

[5] Actually the exchange energy is slightly perturbed which leads to a small change of the Curie temperature (Ref. [4]).

The electric control of magnetic anisotropy affects the domain wall energy. In its turn, it may affect the coercive force, too [7]. Recently the electric control of the coercive force was demonstrated experimentally [31] in InMnAs/InAs structure with perpendicular anisotropy. It is difficult to understand the fivefold change in coercive force without any change in the saturation **M** value by a small (a few percent) change of both $T_c$ and hole density inside InMnAs. On the contrary, this result can be explained by the coupling of FM and SC: the bias electrode controls magnetic anisotropy energy $E_{exc}$ and, in its turn, the coercive force.

---

[6] The authors [29] manipulate the hysteresis loop on *one* sample (Fig. 3 of [29]) whereas the dependence $T_c(V_G)$ is measured on a *different* sample (Fig. 4 of [29]).

[7] The first suggestion on electric control of both magnetic anisotropy and coercive force in the FM/SC hybrids was based on magnetopiezoelectric effect [9]. It combines piezoelectric properties of GaAs-type SC and magnetostriction of FM: changing the electric field in the FM/SC interface region changes the stress (piezoelectricity) varying both the *in-plane* magnetic anisotropy (magnetostriction) and coercivity. To our knowledge this effect has not been observed yet. Note, that besides hybrids the magnetopiezoelectric effect can take place in bulk ferromagnets with piezoelectric properties (for example, in GaMnAs or InMnAs ferromagnets having like GaAs no inversion center).

## 4. Semiconductor-assisted magnetization reversal

The experiments and theoretical arguments discussed above strongly suggest that in equilibrium conditions the SC and FM film form a unified strong-coupled spin system whose magnetic properties can be tuned optically or electrically. The next important step is *dynamical* magnetization reversal *with the help of SC*. The physical idea [25] is illustrated in Fig. 4. The equilibrium situation is shown on Fig. 4a (compare with Fig.3). Break off suddenly the coupling of FM and SC, i.e. put $H_h = 0$. Then magnetic moment **M** rotates coherently (Fig. 4b) about demagnetization field with cycling frequency $\omega = \gamma H_D$ and precession period $T = 2\pi/\omega = 1/2\gamma M_z$. Half-period later the in-plane component of **M** is inverted. Recovery of the coupling at this moment brings the system into another stable state with inverted $M_x$ component [see footnote [4] on page 9]. Indeed, if the spin lifetime of holes is short enough then both the hole spin polarization in the quantum well and the exchange field $H_h$ restore fast to previous values and magnetization is fixed as shown on Fig. 4c. One may turn off the field $H_h$ applying the bias pulses $V_G$ to the gate (Fig.3). Electric variation of the normal component of **M** before cycling tunes the precession period (for example, T≈1 ns for $M_z$=25 Oe).

An important feature of the "field-effect-magnetization-reversal" is the absence of huge current densities (~$10^5$-$10^8$ A/cm$^2$) used in half-metallic [32] and metallic [33] systems. Moreover, the requirement of fast spin lifetime of SC system is easily realized at elevated temperatures [10]. While it is a serious demerit in schemes exploiting the non-equilibrium spin [1], this approach turns it into advantage. Non-equilibrium takes place in ferromagnet during time τ=T/2 of magnetization reversal. Hence the relaxation time of **M** to the equilibrium inside of FM itself should be longer than τ. The spin relaxation time in conventional metallic (about 1 ns [34] in Fe, Co, Ni) and new half-metallic (to this date tens of ps [35] has been reported in GaMnAs) ferromagnets, is rather short. Non-metallic ferromagnets look promising for this purpose (for example, the relaxation time of nickel ferrites is up to hundred of nanoseconds [27]).

Magnetization reversal cycle also can be controlled optically by illuminating the hybrid of pulses of circularly polarized light. As previously discussed, it may be related with optical spin orientation of charge carriers in the FM/SC hybrid. Such a process is accompanied by the appearance of an effective (pulsed in this case) exchange magnetic field [21, 22] inducing the coherent rotation of magnetization. Recent experiments [35] have found the rotation of spins of Mn atoms under pulsed excitation of GaMnAs/GaAs. However, further study is necessary to clarify the origin of the observed effect.

## 5. Summary

We presented a vision of FM/SC hybrid as a unified and flexible system. Quite a number of experiments can be interpreted within the strong-coupled spin system of FM and SC. The uniqueness of the unified system consists in its tunability by optical and electrical means. The system constitutes an elementary magnetic storage with semiconductor being not only a substrate but, against the "common sense," an active participant in information processing. An additional degree of freedom consisting in the choice of desirable FM/SC pair among paramagnet semiconductors and a large number of ferromagnetic materials provides many possibilities. Any day now one may anticipate the discovery of new FM/SC hybrids with new examples of semiconductor-assisted optical and electrical control. Dynamic control of ferromagnetism will bring closer the reality of the "marriage" of magnetism and semiconductor electronics. Enthusiasm in this field reminds the 60-s of the last century when scientists looked for the ideal pair of heterogeneous semiconductors to create heterojunction lasers. The present day is in need of an ideal ferromagnet/semiconductor pair, which means no less interesting and actual problem.

Authors greatly appreciate I.A. Merkulov for valuable remarks and critical reading of manuscript, G.V. Astakhov for discussions and providing us unpublished data on GaMnAs transmission spectra, R.I. Dzhioev for discussions and A.V. Koudinov for critical reading of manuscript, M.V. Lazarev for the help in manuscript preparation. The paper is supported in part by RFBR and CRDF grants, Russian Science Support Foundation and programs of Russian Academy of Sciences.

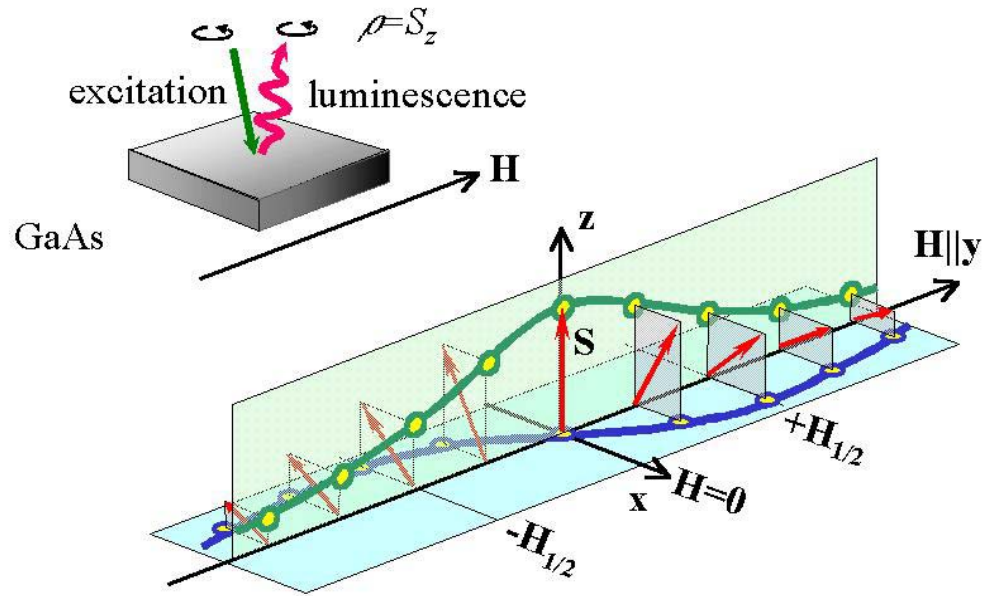

Fig. 1. Geometry of optical orientation experiment (at the top) and the Hanle effect (at the bottom) of semiconductor electrons. Red arrows show the steady state ensemble-averaged electron spin **S** in transversal magnetic field **H** of different values. The projection $S_z$ of electron spin onto observation direction z versus magnetic field is measured (yellow points on the green curve).

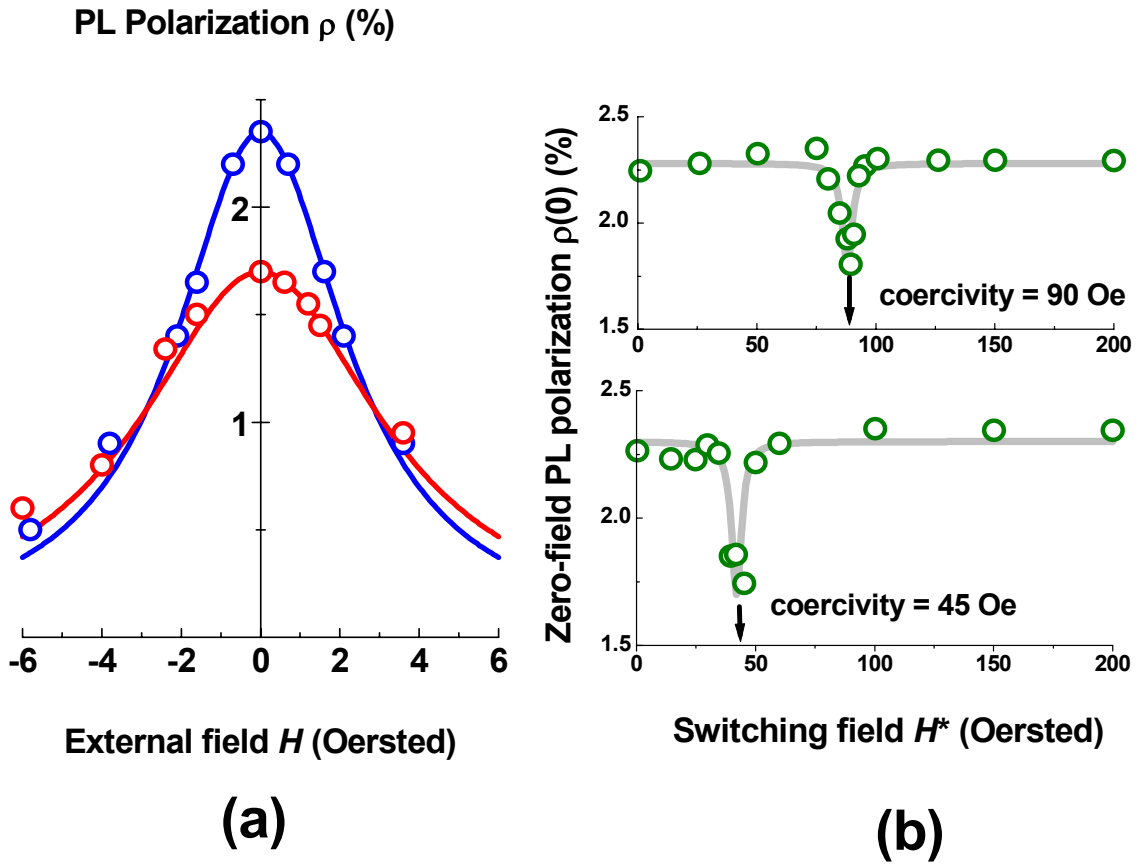

Fig. 2. (a) The Hanle effect in a Ni/n-GaAs structure that was previously demagnetized (red color) and magnetized in plane by the 400 Oe magnetic field (blue color). T=4.2 K.

(b) The degree of zero-field value of circular polarization $\rho(0)$ versus switching field $H^*$. Upper dependence was measured under switching in the dark, whereas the lower one was obtained under illumination during switching process by He-Ne laser with power density 5 W/cm$^2$. (Adapted from [9]).

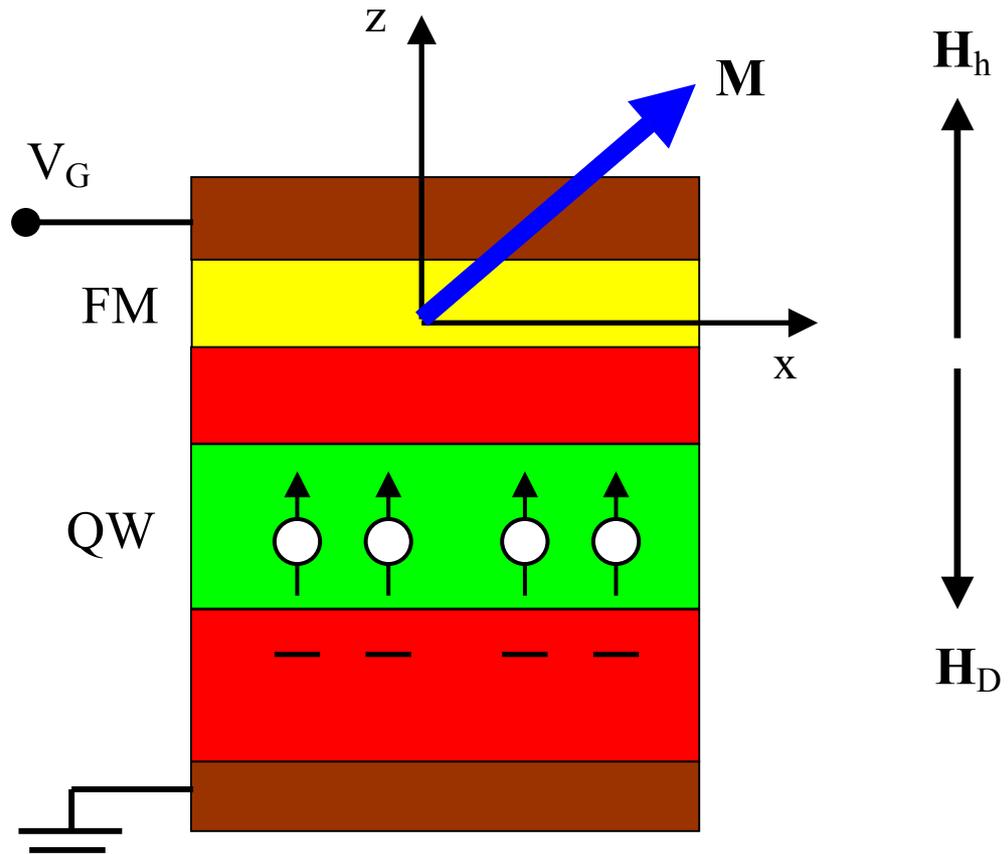

Fig. 3. Cartoon of structure working on exchange coupling of FM and SC. Semitransparent barrier (red layer) separates FM (yellow layer) from quantum well (green layer). Gate (brown layer) bias controls the exchange coupling of FM and holes (open circles) in the well. Arrows denote the spin orientation of holes. (Adapted from [25]).

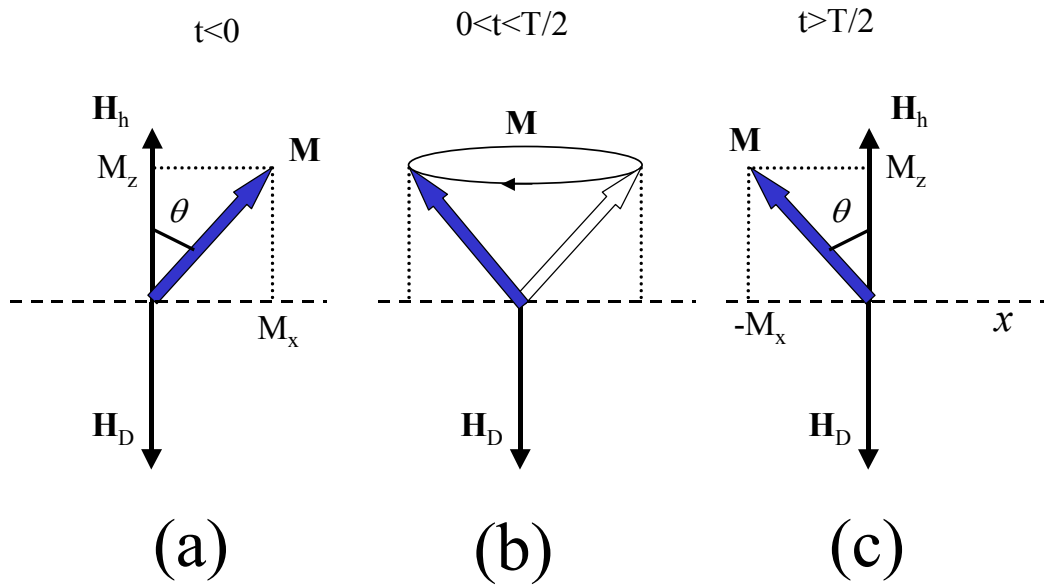

Fig. 4. Reversal of in-plane magnetization component under pulsed turning off the field $\mathbf{H_h}$:

(a)      Stable equilibrium before pulse (t<0). Orientation of magnetization $M_x>0$ in the plane of structure is fixed along easy magnetization axis *x* [footnote [4] on page 9].

(b)      The field $\mathbf{H_h}$ jumps to zero. In the course of half-period [0,T/2] magnetization $\mathbf{M}$ precesses about demagnetization field $\mathbf{H}_D$.

(c)      On completion of pulse (the $\mathbf{H_h}$ field is equal to the datum value) the system is in the equilibrium state with the inverted value of $M_x$.